\documentstyle[preprint,revtex]{aps}

\begin{document}

\hskip.5in \raise.1in\hbox{\fortssbx University of Wisconsin - Madison}
\hfill$\vcenter{\hbox{\bf MAD/PH/743}
                \hbox{February 1993}}$

\vspace{.25in}
\begin{title}
\large \bf
Measuring the Gamma-Gamma Coupling of the Higgs at Linear Colliders
\end{title}
\author{O.\ J.\ P.\ \'Eboli$^1$, M.\ C.\ Gonzalez-Garcia$^2$,
F.\ Halzen$^2$ and\\  D.\ Zeppenfeld$^2$}
\begin{instit}
$^1$ Instituto de F\'{\i}sica,  Universidade
de S\~ao Paulo, \\
Caixa Postal 20516,  CEP 01498 S\~ao Paulo, Brazil.  \\
$^2$ Physics Department, University of Wisconsin, Madison, WI 53706,  USA \\
\end{instit}
\thispagestyle{empty}

\begin{abstract}
Observing the production of the Higgs particle in the $\gamma$-$\gamma$ mode
of a linear $e^+e^-$ collider allows for the measurement of the
$H\gamma\gamma$ coupling. We point out that for the intermediate Higgs mass
range this measurement is considerably more challenging than previously
believed. The $b \bar b$ signature receives a large background from the
production of heavy quark pairs by resolved photons. We quantify the
experimental requirements needed to make a meaningful measurement in the
presence of this background.

\end{abstract}


\newpage

Experiments at future linear $e^+e^-$ colliders, such as the 500 GeV NLC, will
be able to do detailed investigations of the interactions of gauge bosons,
fermions and scalars. One of the prime targets is the detailed study of the
interactions of the Higgs boson, assuming that it is light enough to be
produced at these machines. In particular it has been suggested~\cite{gunhab}
to use the $\gamma\gamma$ collision mode of such a machine, with the photons
generated by backscattered laser beams \cite{las0}, in order to investigate
the production process
\begin{equation}
       \gamma\gamma \rightarrow H
\end{equation}
and thereby achieving a direct measurement of the $H\gamma\gamma$ coupling.
Measuring this coupling should be a high priority as it is in principle
sensitive to all charged particles which acquire their mass via the Higgs
mechanism. The accuracy of this measurement critically depends on the magnitude
of the competing backgrounds. In the analysis of Ref.~\cite{gunhab} the
$\gamma\gamma\rightarrow b\bar b$ background was evaluated and the possibility
of highly polarized photon beams was explored as a way to enhance the signal
over background. In addition a measurement of the $b\bar b$ invariant mass with
an optimistic error of $\pm 2.5$~GeV was assumed.

We here point out that, unfortunately, at a photon collider large additional
irreducible backgrounds will be present which can be traced to the hadronic
content of the photon. The splitting sequence $\gamma \rightarrow  {\rm quark}
\rightarrow  {\rm gluon}$ will provide for an effective beam of strongly
interacting partons with the resultant cross sections for producing $b\bar b$
pairs of strong interaction size \cite{thres,Drees}. As a result additional
irreducible backgrounds arise from the processes
\begin{equation}
\begin{array}{l}
   \gamma g \rightarrow b \bar b \\
      g g   \rightarrow b \bar b \; ,
      q \bar q   \rightarrow b \bar b \; ,
\end{array}
\end{equation}
and corresponding backgrounds appear due to the even larger production cross
sections of charmed quarks. The latter represent, in principle, a reducible
background although they have the potential to fake b-quarks in a realistic
experiment as they exceed the b-pair background by up to one order of
magnitude.
One additional problem is that the polarization information of the original
photon is largely lost in the splitting $\gamma \rightarrow  {\rm gluon}$ and
therefore the use of polarized photon beams will be ineffective in reducing
this resolved photon background.

A defining feature of the $H\rightarrow  b\bar b$ signal is the S-wave nature
of the resonance which results in the production of the $b$-quarks at large
c.m. scattering angles. In contrast all the backgrounds peak at small
scattering angles. Following Ref.~\cite{gunhab} we therefore require that the
scattering angle of the b's with respect to the beam axis is large enough,
$|cos\theta|<0.85$ in the $b\bar b$ rest frame. Imposing this cut in all of
the following, we compare in Fig. \ref{sigmas}.a the total cross section of
the Higgs signal $\sigma (H\rightarrow  b\bar b)$, measured in fb, with the
invariant mass  distributions (measured in fb/10~GeV) of the various
backgrounds.  For the $\gamma\gamma\rightarrow  b\bar b$ background the
reduction due to polarized beams with $<\lambda \lambda^\prime> = 80$\% has
been included. It amounts to a factor of 5. Convolution with realistic photon
distributions
\cite{lasernew,thres}
is already  included in all cross sections, therefore the corresponding number
of events will be given by  $N_{ev}={\cal L}_{e^+e^-} k \sigma$ where the
conversion coefficient $k$ represents the average number of high energy photons
per electron. For the 500 GeV NLC  a reasonable assumption is ${\cal
L}_{e^+e^-} k=10$-$100$ fb$^{-1}$ year$^{-1}$ \cite{lasernew,pal}.

One finds that for all interesting Higgs masses the resolved photon background
is at least a factor 3--8 larger than the  $\gamma\gamma\rightarrow  b\bar b$
background and this factor approaches 2 orders of magnitude for Higgs masses
around 70~GeV.    The dominant resolved photon background is dominated by the
process $\gamma g \rightarrow  b \bar b$.  Gluon-gluon fusion and $q$-$\bar q$
annihilation are, in comparison, negligeable. For the results  shown in Fig.
\ref{sigmas}.a we have summed all resolved contributions. In the  $\gamma g$
process the photon typically carries much more energy in the laboratory frame
than the gluon which leads to a substantial boost of the event along the beam
axis in the incoming photon direction. As a result the larger of the two
b-rapidities is bigger in the resolved photon background than in the signal as
is shown in Fig. \ref{sigmas}.b. As illustrated in table \ref{signific},
requiring a maximum value for the b-quark jet rapidities leads to a
substantial improvement in the signal to background ratio. As seen from the
table, the improvement is optimal for $|y_b|\raisebox{-.4ex}{\rlap{$\sim$}}
\raisebox{.4ex}{$<$}1.5$, a value which compares favorably with realistic
coverage assumptions of the microvertex detectors which will be essential for
$b$-identification.

Because the gluon content of the photon has not been reliably measured to date
(experiments at HERA should soon remedy this situation), there is considerable
uncertainty regarding the effective $b\bar b$ production cross section. We
have used two presently popular sets of gluon distribution functions inside
the photon to span these uncertainties, the parameterization by Drees and
Grassie~\cite{DG} (DG) which provides for a relatively soft gluon distribution
and the LAC3 parameterization of Ref.~\cite{LAC3} which gives a considerably
harder gluon distribution. Use of the LAC3 structure function not only
results in significantly larger resolved backgrounds, but the difference in
rapidity distribution between signal and background, shown in Fig.
\ref{sigmas}.b, is further reduced. Because of the presence of harder gluons
the photon-gluon CM-frame receives a smaller boost along the beam axis and the
$b$-rapidity distribution overlaps even more with the one for the Higgs signal.

Because the $\gamma\gamma\rightarrow  q\bar q$,  $\gamma g\rightarrow  q\bar
q$, and  $gg \rightarrow  q\bar q$ cross sections scale like $Q_q^4$, $Q_q^2$
and $Q_q^0$, respectively, where $Q_q$ denotes the quark electric charge, the
charm production cross sections are larger by a factor 16 and 4 respectively
for $\gamma\gamma$ and $\gamma g$ fusion. Hence an excellent suppression of
charm events is required. At the same time an excellent $b\bar b$ invariant
mass resolution is required in order to identify the $H\rightarrow  b\bar b$
invariant mass peak. Unfortunately these two requirements are at least
partially exclusive: leptonic $B$-decay would provide a powerful tool in
suppressing the charm background, however, every leptonic decay in the
$b\rightarrow  c \rightarrow $ light quarks decay chain produces missing
momentum due to escaping neutrinos and hence intrinsically limits the mass
resolution. Excellent mass resolution is only possible when restricting to
events without leptonic decays, with double displaced vertices as a signature
for both a $B$ and a subsequent $D$ decay. While this imposes severe
requirements on the performance of the micro-vertex detectors, it also
eliminates about 70\% of all signal events. The only practical solution appears
to be to accept leptonic decay modes and hence a reduced mass resolution. The
problem has been considered before in Ref.~\cite{grindh} in the context of
Higgs production at LEP/LHC. The intrinsic mass resolution of the $H
\rightarrow b\bar b$ peak was found to be around $\pm 10$~GeV for a 120~GeV
Higgs, with substantial tails due to missing particles. Any improvement come at
the price of substantial loss in rate. Somewhat optimistically we assume in the
following that the Higgs peak will be contained in a 20~GeV bin centered at
$m_H$ and we integrate the backgrounds over the same bin when making rate
comparisons.

The results are shown in Fig. \ref{sigmascut} where we have ignored any charm
background. Even for a soft gluon distribution inside the photon the Higgs
signal never is larger than half the background cross section. The total
background is strongly dependent on the rejection of charm events. A 90\%
efficiency for rejecting charm, combined with a 90\% identification
probability of actual $b\bar b$ events would lead to a value of
$\sigma_{\mbox{signal}}/\sqrt{\sigma_{\mbox{background}}} = 2.2$~fb$^{1/2}$
for $m_H=140$~GeV and using DG structure functions, which for an integrated
luminosity of 10~fb$^{-1}$ corresponds to a 7 sigma signal. Obviously this
result is based on quite optimistic assumptions and the Higgs mass range for
which the result applies is only of order 20 GeV. For lighter masses the
situation worsens rapidly, as is evident from Table~I. It should be pointed
out that even a  7$ \sigma$ result does not allow a determination of the top
quark loop contribution to the $H\gamma\gamma$ coupling.

A last way to reduce the background is by tagging the resolved
photon by its beam jet. The efficiency of this is completely dictated by the
achitecture of the detector. Obviously good coverage at small angles is
required. We summarize our conclusions by emphasizing once more that in order
to measure the $H\gamma\gamma$ coupling over most of the intermediate mass
range (60--140~GeV) one requires a detector with i) good $b$-identification
in the central region ii) excellent charm rejection and iii) identification of
narrow angle beam jets associated with resolved photons. We also argued that
improving the situation by assuming $b\bar b$ mass resolutions as small as $\pm
2.5$~GeV~\cite{gunhab} is quite unrealistic.

\acknowledgments
This work was supported by the University of
Wisconsin Research Committee with funds granted by the Wisconsin Alumni
Research Foundation, by the U.S.~Department of Energy under Contract
No.~DE-AC02-76ER00881, by the Texas National Research Laboratory Commission
under Grant No.~RGFY9273.

\figure{{\bf (a)} Total cross section for the Higgs signal
$\sigma (H\rightarrow  b\bar b)$  measured in fb (solid line) and invariant
mass distributions (measured in fb/10~GeV) of the various backgrounds.
The dashed lines correspond to the direct photon backgrounds.
Dotted (dash-dotted) lines correspond to resolved photon backgrounds for
DG  (LAC3) photon structure functions. In all cases the lower (upper) line
correspond to $b\bar b$ ($c\bar c$)  background.
{\bf (b)} Maximum rapidity distribution of $b$ quarks with invariant mass
$m_{bb}=130\pm 10$ GeV. The dotted histogram correspond to $b$'s from
the decay of a Higss with $m_H=130$ GeV. The solid lines correspond to
the $\gamma\gamma\rightarrow b\bar b$ background  (lower) and
$\gamma g\rightarrow b\bar b$ (upper) for the DG photon structure functions.
In all cases we require $|cos\theta|_Q<0.85$ in the $Q\bar Q$ rest frame.
For the $\gamma\gamma\rightarrow  Q\bar Q $ background the reduction due to
polarized beams with $<\lambda \lambda^\prime> = 80$\% has been included.
\label{sigmas}}
\figure{Cross section for the Higgs signal and backgrounds (same notation as
Fig. 1). For the backgrounds an invariant mass resolution $m_H\pm 10$~GeV is
assumed. In all cases we require $|cos\theta|_Q<0.85$ in the $Q\bar Q$ rest
frame and a maximum rapidity $|y_{max}|_Q<1.5$.
For the $\gamma\gamma\rightarrow  Q\bar Q $ background the reduction due to
polarized beams with $<\lambda \lambda^\prime> = 80$\% has been included.
\label{sigmascut}}

\begin{table}
\begin{center}
\begin{displaymath}
\begin{array}{||c|c|c|c|c|c||}
\hline
\hline
\multicolumn{6}{||c||}
{\displaystyle {\sigma_{\mbox{signal}}/\sqrt{\sigma_{\mbox{background}}}
(\sqrt{fb})} }\\
\hline
M_{\mbox{Higgs}} {\mbox{(GeV)}} & y_{\mbox{max}}
& \gamma\gamma\rightarrow b \bar b   &
\gamma\gamma\rightarrow c \bar c    & \gamma g\rightarrow b \bar b    &
\gamma g\rightarrow c \bar c \\
\hline
\hline
     &\infty & 1.2   & 0.35 & <0.14 & <0.007 \\
60   &   2      & 1.1   & 0.32 & <0.18 & <0.09  \\
     &   1.5    & 0.98  & 0.29 & <0.21  & <0.11  \\
     &   1.     & 0.77  & 0.22 & <0.2  & < 0.1  \\
\hline
     &\infty & 7.2   & 1.86 & <3.1 & <1.55   \\
 140 &   2      & 7.9   & 1.94 & <3.3 & <1.65    \\
     &   1.5    & 7.3   & 1.88 & <3.5 & <1.76   \\
     &   1      & 5.6   & 1.45 & <3.2 & <1.65   \\
\hline
\hline
\end{array}
\end{displaymath}
\end{center}
\caption{ ${\sigma_{\mbox{signal}}/\sqrt{\sigma_{\mbox{background}}}
(\sqrt{fb})}$ for the main backgrounds. For the backgrounds an invariant mass
resolution $m_H\pm 10$~GeV is assumed.For the resolved photon backgrounds the
numbers correspond to DG structure functions. The significance of the signal
in standard deviations is obtained multiplying this numbers by
$\sqrt{{\cal{L}}_{e^+e^-} k} \approx 3$-$10$ fb$^{-1/2}$.} \label{signific}
\end{table}

\end{document}